\documentstyle[12pt,epsf]{article}
\begin{document}
\thispagestyle{empty}
\noindent
{\bf Preprint FZR-188 (1997)}\\
Phys. Lett. B, in print

\vspace*{2cm}

\begin{center}
{\large \bf
Estimates of dilepton spectra from open charm and bottom decays
in relativistic heavy-ion collisions}\\[1cm]
{\sc B. K\"ampfer$^1$,
O.P. Pavlenko$^{1,2}$,
K. Gallmeister$^1$}\\[1cm]
$^1$Research Center Rossendorf, PF 510119, 01314 Dresden, Germany\\
$^2$Institute for Theoretical Physics, 252143 Kiev, Ukraine
\end{center}

\vspace*{2cm}

\centerline{Abstract}
The spectra of lepton pairs from correlated
open charm and bottom decays in ultrarelativistic heavy-ion collisions
are calculated. Our approach includes energy loss effects of heavy quarks
in deconfined matter
which are determined by temperature and density of the expanding parton
medium. We find a noticeable suppression of the initial transverse momentum
spectrum of heavy quarks due to the energy loss at LHC conditions.
Within the central rapidity covered by the ALICE detector system, the dominant
contribution from bottom decays to the high invariant mass dielectron
spectrum is predicted.\\[1cm]

\newpage
\setcounter{page}{1}


The production of high invariant mass dileptons in relativistic
heavy-ion collisions has attracted a great deal of interest in the recent years
from both the experimental and theoretical sides.
Dilepton measurements are planned
with the PHENIX and ALICE detector facilities at the
relativistic heavy-ion collider RHIC in Brookhaven and at the large hadron
collider LHC in CERN, respectively.
In despite of many theoretical attempts
the nature of the dominating source of dileptons,
which will be measured in these detectors, is still unclear.
Among the candidates that can give
strong contributions in the large invariant mass continuum region near and
beyond the $J/\psi$ one can consider the following processes in which
dileptons are generated:
(i) hard initial quark - anti-quark collisions
give rise to the Drell-Yan yield,
(ii) semi-leptonic decays of heavy quarks like the charm and bottom,
which are also produced dominantly in hard initial parton interactions,
and
(iii)
the thermal dilepton radiation resulting in
interactions between secondary partons or even in a
locally thermalized quark-gluon plasma (QGP).
The yield from (iii) is expected to serve as a direct probe from
the QGP \cite{Shur3,Ruusk}.

It is commonly believed that, due to the large invariant mass $M$, at least
the Drell-Yan process and the initial heavy quark production in pp and
pA collisions are under reliable theoretical control by means of
perturbative QCD \cite{prob1,prob2,Sar}
and, therefore, can provide a reference for the thermal dilepton
signal from deconfined matter. At the same time all up to date calculations
of high-$M$ thermal dilepton rates from early parton matter contain
uncertainties due to the lack of knowledge of precise initial condition
parameters such as the energy density and quark-gluon phase space saturation.
The parton cascade model \cite{Geiger} and calculations based on the
HIJING event generator \cite{Wang} differ substantially
in predicting the thermal high-$M$ dilepton spectra. Recent estimates
\cite{our_PL} within the mini-jet mechanism of quark-gluon matter formation
\cite{Eskola}, with initial conditions similar to the
self-screened parton cascade model \cite{E.M.W.}, point to a strong
competition between the Drell-Yan yield and the thermal signal in a wide
range of beam energies of the colliding nuclei.

With respect to the wanted dilepton signal from the QGP
the dileptons from charm and bottom decays play a r\^ole of
substantial background \cite{Vogt2}.
The magnitude and the characteristic properties of such background is
presently matter of debate. As shown by Shuryak \cite{Shur1}
for RHIC conditions
the energy loss effects of heavy quarks propagating through a medium
can drastically reduce the dilepton yield from charm and bottom decays
in the region of large invariant dilepton masses.
Further calculations \cite{Lin2}, including the PHENIX acceptance
at RHIC, also predict a very strong suppression of dileptons from heavy
quark decays, which then is partially below the Drell-Yan yield.
The studies in refs.~\cite{Shur1,Lin2} are based on the
approximation of a constant rate of energy loss
$- dE/dx =$ 1 or 2 GeV/fm. Independently of the assumed mechanism of
energy loss the approximation of a constant value of $dE/dx$
is difficult to justify for strongly expanding and cooling matter.
In general, the value of $dE/dx$ is expected to depend not only on the
geometry and a fixed mean free path of heavy quarks
but also on the temporal evolution of the medium.
As shown below, in particular within the gluon radiation mechanism
of energy loss \cite{Baier,Baier2,Zakharov,Baier_new},
$dE/dx$ can depend on widely varying characteristic quantities
of the parton medium formed at RHIC and LHC.

In the present paper we study the dileptons resulting
from decays of correlated open charm and bottom
produced initially in hard collisions and undergoing an energy loss
which is determined by the temperature and the density evolution
of the deconfined medium. Going beyond refs.\ \cite{Shur1,Lin2} we
concentrate here on LHC conditions, where the
energy loss appears to be stronger as compared to RHIC.
In a certain scenario
this leads to almost complete stopping of
initial charm and bottom in deconfined matter produced at LHC.
The central rapidity acceptance of ALICE will select
in this case the electron pairs mainly from bottom decays.
It is assumed that, due to a like-sign subtraction, the combinatoric
background from uncorrelated pairs can be removed,
and henceforth we focus on correlated pairs.


In the  framework of perturbative QCD the number of the heavy
quark - anti-quark ($Q \bar Q$) pairs, produced
initially with transverse momenta $p_{\perp 1} = -p_{\perp 2} =  p_\perp$
at rapidities $y_1$ and $y_2$, in
central AA collisions can be calculated from
\begin{equation} \label{eq.1}
dN_{Q \bar Q}= T_{AA}(0) \, {\cal K}_Q \, H(y_1,y_2,p_\perp) \,
dp_\perp^2 \, dy_1 \, dy_2,
\end{equation}
where $H(y_1,y_2,p_\perp)$ is the standard combination
of structure functions and elementary cross sections
(see for details \cite{our_PL,Vogt2})
\begin{eqnarray}
H(y_1,y_2,p_\perp)
& = &
x_1 \, x_2 \left\{ f_g(x_1,\hat Q^2) \, f_g(x_2,\hat Q^2)
\frac{d \hat \sigma_g^Q}{d \hat t} \right. \nonumber \\
& + &
\left. \sum_{q \bar q} \left[
f_q(x_1,\hat Q^2) \, f_{\bar q} (x_2,\hat Q^2) +
f_q(x_2,\hat Q^2) \, f_{\bar q} (x_1,\hat Q^2)\right]
\frac{d \hat \sigma_q^Q}{d \hat t} \right\},
\end{eqnarray}
where $f_i(x,\hat Q^2)$ with $i=g,q,\bar q$
are the parton structure functions,
$x_{1,2} = m_\perp \left(
\exp\{ \pm y_1 \} + \exp\{ \pm y_2 \}  \right) /\sqrt{s}$ and
$m_\perp = \sqrt{p_\perp^2 + m_Q^2}$.
As heavy quark masses we take $m_c =$ 1.5 GeV and $m_b =$ 4.5 GeV.
We employ throughout the present paper
the HERA supported structure function set MRS D'- \cite{MRS} from the PDFLIB
at CERN. The overlap function for central collisions is
$T_{AA}(0) = A^2/(\pi R_A^2)$
with  $R_A = 1.1 A^{1/3}$ fm and $A = 200$ in this paper.
We here do not include shadowing effects of nuclear parton distributions
since for heavy quark production
they are expected to be not very important
and can be considered separately.
Our calculation procedure is based on the lowest-order QCD
cross sections $d \hat \sigma_{q,g}^Q / d\hat t$
for the subprocesses $gg \to Q \bar Q$ and $q \bar q \to Q \bar Q$
with the simulation of higher order corrections by the corresponding
${\cal K}_Q$ factor.
Such a procedure reproduces within the needed
accuracy the next-to-leading order calculations \cite{Vogt2}
of the heavy quark pair
distributions with respect to their invariant mass,
total pair rapidity and rapidity gap. We find the scale $\hat Q^2=4m_Q^2$
and ${\cal K}_Q=2$ as most appropriate.

For the Drell-Yan production process of leptons at rapidities $y_1$ and
$y_2$ and transverse momenta $p_{\perp 1} = - p_{\perp 2} = p_\perp$
we have
\begin{eqnarray} \label{eq.1l}
dN_{l \bar l}^{DY} & = & T_{AA}(0) \, {\cal K}_{DY} \, L(y_1,y_2,p_\perp) \,
dp_\perp^2 \, dy_1 \, dy_2, \\
L(y_1,y_2,p_\perp) & = & \sum_{q,\bar q} x_1 x_2
\left[
f_q(x_1,\hat Q^2) \, f_{\bar q} (x_2,\hat Q^2) +
f_q(x_2,\hat Q^2) \, f_{\bar q} (x_1,\hat Q^2) \right]
\frac{d \hat \sigma_q^{l \bar l}}{d \hat t},
\end{eqnarray}
with
$d \hat \sigma_q^{l \bar l} / d \hat t = \frac{\pi \alpha^2}{3}
\mbox{ch} (y_1 - y_2) / \left( 2 p_\perp^4
[1 + \mbox{ch}(y_1 - y_2)]^3 \right)$,
$x_{1,2} = p_\perp \left(
\exp\{ \pm y_1 \} + \exp\{ \pm y_2 \} \right)$ $/\sqrt{s}$,
$\alpha = 1/137$, $\hat Q^2 = x_1 x_2 s = M^2$
and ${\cal K}_{DY} =$ 1.1 \cite{prob1}.


Since the energy loss of heavy quarks depends on the properties
of produced matter, which is propagated through,
one has to specify the initial conditions and the
space-time evolution of such a medium. To do this in a coherent
manner we employ for the mini-jet production (which dominates the parton
matter formation) the same lowest-order
approximation and parton structure functions
as for the initial heavy quark production.
Adding a suitably parametrized soft
component \cite{our_PL} we get the initial temperature
$T_i = 1000$ (550) MeV, gluon fugacity $\lambda_i^g = 0.5$
and light quark fugacity $\lambda_i^q = \lambda_i^g / 5$
of the mini-jet plasma formed at initial time
$\tau_i = 0.2$ fm/c at LHC (RHIC).
The space-time evolution of the produced parton matter
after $\tau_i$ is governed by the longitudinal scaling-invariant expansion
accompanied by
quark and gluon chemical equilibration processes \cite{Biro,our_PRC}.
We take for definiteness full saturation, i.e.
$\lambda^g = \lambda^q = 1$ at deconfinement temperature $T_c = 170$ MeV.
(Due to gluon multiplication \cite{Shur2} this seems to be justified, while
the chemical equilibration of the quark component might be somewhat
overestimated by this ansatz.)
The actual time evolution
is determined by $e \propto \tau^{-4/3}$ for energy density and
$\lambda^{q,g} \propto \tau^2$ \cite{our_PRC}, which must be inverted to get
$T(\tau)$ explicitly.
This results in final pion rapidity densities
(see \cite{our_PL})
which are in agreement with other estimates \cite{Eskola}.

To model the energy loss effects of heavy quarks in expanding matter we assume
as usual in the Bjorken scaling picture that a heavy quark produced initially
at given rapidity will follow the longitudinal flow with the same
rapidity. Therefore, with respect to the fluid's local rest frame
the heavy quark has essentially only a transverse momentum $p_\perp$
which may depend on the proper local
time $\tau$ in accordance with the energy loss in transverse direction.
(Note that the transverse expansion at early times can be neglected,
and we take as averaged transverse radius of parton matter $R = 7$ fm.)

Recently the total energy loss of a high-energy parton propagating
transversally through expanding deconfined matter has been derived
as $dE / dx |_{\rm expanding} = \xi dE / dx |_{T_f}$ \cite{Baier_new},
supposed the medium cools according to a power law (or similar) and the
initial time tends to zero. The numerical factor $\xi$ is of the order
2 (6) for a parton created inside (outside) the medium.
The subscript $T_f$ indicates that the stopping power depends
only on the final temperature at which the deconfined medium is left.
We shall denote hereafter this stopping scenario as model I and
take $dE / dx |_{T_f}$ from \cite{Baier,Baier2,Zakharov,Baier_new}
with the new, improved numerical factors.

We are going to contrast model I with another one which relies on
an accumulative energy loss similar to \cite{Shur1,Lin2}.
To calculate the evolution of the transverse momentum $p_\perp(\tau)$ of
quarks propagating the distance $r_\perp (\tau)$ in transverse direction
we adopt the results of \cite{Baier} for the
energy loss of a fast quark in a hot QCD medium (model II)
\begin{eqnarray}
\label{eq.21}
\frac{d p_\perp}{d \tau} & = & - \frac43
\frac{\alpha_s k_c}{\sqrt{L}} \, (p_\perp^2 + m_Q^2)^{1/4} \,
\ln \left( \frac{\sqrt{p_\perp^2 + m_Q^2}}{L \, k_c^2} \right),\\
\label{eq.22}
\frac{d r_\perp}{d \tau} & = & \frac{p_\perp}{\sqrt{p_\perp^2 + m_Q^2}},
\end{eqnarray}
where the parameter
$k_c^2 = 2 m_{th}^2(T) \equiv \frac 89 (\lambda_g +\frac 12 \lambda_q)
\pi \alpha_s T^2$
\cite{Biro,our_PRC} is used as an average of the
momentum transferred in heavy quark-parton scatterings,
and the strong coupling is described by $\alpha_s = 0.3$.
For the mean free path of gluons we take
$L \approx (\sigma_{gg} n_g)^{-1}$ with integrated cross section
$\sigma_{gg} \sim \alpha_s^2/k_c^2$ and gluon density
$n_g \sim \lambda_g T^3$,
resulting in $L^{-1} = 2.2 \, \alpha_s \, T$
\cite{Biro,Shur2}.
Strictly speaking, eqs.~(5,\,6) are valid for a static medium
under certain conditions \cite{Baier2}. Integration of these equations
therefore neglects the destructive interferences of the
Landau-Pomeranchuk-Migdal effect.
The r.h.s. of eq.~(\ref{eq.21}) is actually the rate of energy loss $dE/dx$
which is essentially $\propto (p_\perp^2 + m_Q^2)^{1/4}$.
Due to the expansion of the matter $dE/dx$
depends on the time.
For instance, at LHC conditions and initial momentum of a heavy quark of
10 GeV the stopping power derived from eq.~(5)
drops from 6 GeV/fm down to 0.3 GeV/fm
during the expansion of deconfined matter.

To get the spectra of charm and bottom quarks
after energy loss we use a Monte Carlo
simulation with a uniform distribution of the random initial position
and random orientation of $Q \bar Q$
pairs in the transverse plane.
In model I the total energy loss is determined by the transverse distance
$d$ of a created heavy quark to the boundary of the system, the quark
initial energy and the temperature when leaving the system.
Most quarks experience an energy loss $\propto d$ and only a few ones
$\propto d\,^2$.

In model II we integrate eqs.~(\ref{eq.21},\ref{eq.22})
together with the above described time evolution of
$T(\tau)$ and $\lambda_{q,g}(\tau)$. Eq.~(\ref{eq.22}) is used to check
whether the considered heavy quark propagates still within
deconfined matter; if it leaves the deconfined medium it does not longer
experience the energy loss according to eq.~(\ref{eq.21}).
(The subsequent energy loss in a mixed and hadron phase might
be incorporated along the approach of \cite{Sv}.)
A considerable part of heavy quarks can not
escape the parton system thereby undergoing thermalization.
Following \cite{Lin2}
we consider a heavy quark to be thermalized if its transverse mass
$m_\perp$
during energy loss becomes less than the averaged transverse mass
of thermalized light partons at given temperature.
These heavy quarks we assume to be distributed as
$dN_Q / dm_\perp \propto m_\perp^2 \exp\{ -m_\perp / T_{\rm eff} \}$
with an effective temperature $T_{\rm eff} = 150$ MeV.

The results of our calculation for
transverse momentum spectra of charm and bottom quarks
at midrapidity are shown in fig.~1. Within model II
we obtain a strong suppression of the initial transverse momentum
spectrum at large transverse momenta due to the energy loss both for charm and
bottom. The bump below 3 GeV is caused by the thermal redistribution.
In comparison with RHIC conditions \cite{Lin2,our_cont}
the suppression at LHC is found to be more pronounced due to the longer
life time of deconfined matter and its higher initial temperature and density.
It should be stressed that such a drastic change of the $p_\perp$ spectrum
of heavy quarks
in AA collisions, in comparison with pp or pA collisions,
would be a measurable effect which for itself can help to
identify the creation of a hot and dense parton gas \cite{Wang,Tannenbaum}.
In contrast to model II,
model I results in a smaller depletion of the large-$p_\perp$
spectrum (see fig.~1).


Basing on the analysis of ref.\ \cite{Vogt2} we employ a delta function like
fragmentation function for heavy quark fragmentation into D and B mesons.
It gives the same transverse momentum for the meson as for the parent
heavy quark.
The correlated lepton
pairs from $D \bar D$ and $B \bar B$ decays have been obtained from a suitable
Monte Carlo code which employs the single electron distributions
as delivered by JETSET 7.4.
The average branching ratio of $D \to l + X$ is taken to be 12\%.
We consider here only the so-called primary leptons \cite{Lin2}
which are directly produced in the bottom decay
$b \to l \, X$ with the average branching ratio 10\%.
The secondary leptons $l'$ produced in further branchings like
$b \to c \, X \to l' Y$ have much smaller energy and do not contribute
to the large invariant mass region. Such simplification does not affect
significantly the dielectron yield at $M > 2$ GeV as we have checked by
comparison with ref.\ \cite{Vogt2}.
To get the dielectron spectra from charm and bottom
decays at LHC we take into account the acceptance of the ALICE detector:
the electron pseudo-rapidity is $|\eta_e| < 0.9$ \cite{ALICE_prop}
with an additional cut on the electron transverse momentum
$p_\perp^e > 1$ GeV
to reduce the lepton background from light hadron decays.
We do not take into account the limited high-$p_\perp$ acceptance
($p_\perp^e \le 2.5$ GeV \cite{ALICE_prop})
in our simulations to demonstrate the
general tendency of the spectrum in the high-$M$ region.
Our results of heavy quark decays at LHC are depicted in fig.~2 for charm
(a) and bottom (b) for $M \ge 2$ GeV. Due to the strong energy loss
in model II
the contribution from charm decays drops below the Drell-Yan yield.
At the same time the contribution
from bottom decays becomes dominant up to $M =$ 6 GeV.
This can be
understood also on a qualitative level. Even after strong energy loss
the massive B mesons have enough energy to produce electrons with
$p_\perp^e > 1$ GeV. Additionally the central rapidity cut of the ALICE
detector selects electrons from charm being near mid-rapidity,
while for bottom
the corresponding rapidity region is greater.
Further, we actually estimate a lower limit for bottom decay;
as discussed above the secondary electrons contribute somewhat
at not too large values of $M$.
Therefore, at LHC the dielectron continuum region above 2 up to
6 GeV is expected to be dominated by bottom decays.
This prediction is also supported by model I.
Here the energy loss is smaller, and the lepton pairs
from bottom dominate at larger invariant mass.
At $M <$ 5 GeV bottom and charm are equally strong, however
an additional energy shift, e.g. caused by
including a non-$\delta$ like fragmentation, lets bottom win.
(The Peterson fragmentation function results in a $>3$ times stronger bottom
contribution in comparison with charm. Also a slightly changed numerical
factor, say $\xi =$ 3, suppresses the charm contribution down to
the Drell-Yan yield, while the bottom contribution is less affected).

To demonstrate the general tendency of the relative contribution from the
Drell-Yan process and heavy quark decays with the change of the collider
energy we plot in fig.~3 the results of our calculations for RHIC initial
conditions including the PHENIX detector acceptance for the
electrons, i.e. $|\eta_e| \le 0.35$ \cite{PHENIX_prop}.
As above for LHC we here also employ
the gate $p_\perp^e > 1$ GeV. For model II
one observes in fig.~3 a strong competition
of the Drell-Yan process and correlated heavy quark decays in a wide region
above $M = 2$ GeV similar to \cite{Lin2}.
Imposing an additional kinematical cut, which selects back-to-back
pairs, Drell-Yan dileptons very clearly dominate. However, notice that
model I does not give a noticeable effect of energy loss at RHIC
conditions.


In summary, taking into account the energy loss effects of heavy
quarks we calculate the spectra of electron pairs
from correlated open charm and bottom decays at LHC conditions
within two different scenarios.
Within the naive scenario with accumulative energy loss
in the expanding deconfined matter
we obtain a measurable, strong suppression of large transverse momenta
of heavy quarks,
while the recent treatment of the energy loss in \cite{Baier_new}
predicts a smaller effect.
With respect to these different results and uncertainties inherent
in our approach (like the neglect of quark masses in some energy loss
formulae, finite energies of quarks, use of leading order results
in the strong coupling constant, neglect of the confinement transition
and the influence of confined matter) any prediction seems
at best semi-quantitative. Therefore, an experimental determination of the
transverse momentum spectra of heavy quarks in future heavy-ion
collisions appears as challenging task to pin down the essential effects.
Qualitatively,
our calculations predict the dominating contribution of bottom decays
in the high invariant mass region of electron pairs which are planned
to be measured at central rapidity in the ALICE detector.

Useful discussions with H.W. Barz, Yu.L. Dokshitzer, Z. Lin, A.H. Mueller,
E.V. Shuryak, R. Vogt, X.-N. Wang, and G.M. Zinovjev
are gratefully acknowledged.
We are indebted to R. Baier for stimulating discussions and for
informing us on his results prior to publication.
O.P.P. thanks for the warm hospitality of the nuclear theory group
in the Research Center Rossendorf.
The work is supported by BMBF grant 06DR829/1.

\newpage

\centerline{{\bf Figure captions}}

\vspace*{6mm}

{\bf Fig.~1:}
The $p_\perp$ spectrum of heavy quarks ((a): charm, (b): bottom)
at midrapidity
in AA collisions at LHC ($\sqrt{s} = 5500 $ AGeV). The dotted curves depict
the initial production without energy loss,
while the dashed and solid lines show our result
with energy loss according to model I and II
and $T_{\rm eff} = 150$ MeV for thermalized heavy quarks.

\vspace*{6mm}

{\bf Fig.~2:}
The invariant mass spectrum of dielectrons from correlated charm (a) and
bottom (b) decays as well as Drell-Yan electron pairs
filtered throughout the LHC-ALICE acceptance,
but the high-$p_\perp$ cut of
2.5 GeV is here not implemented.

\vspace*{6mm}

{\bf Fig.~3:}
The invariant mass spectrum of dielectrons from correlated charm (dotted line)
and bottom (solid line) decays
as well as Drell-Yan electron pairs (dashed line)
filtered throughout the PHENIX acceptance. The energy loss is calculated
within model II.

\newpage

\begin{figure}[h]
\centering
~\\[.1cm]
\centerline{\epsfxsize=.7 \hsize \epsffile{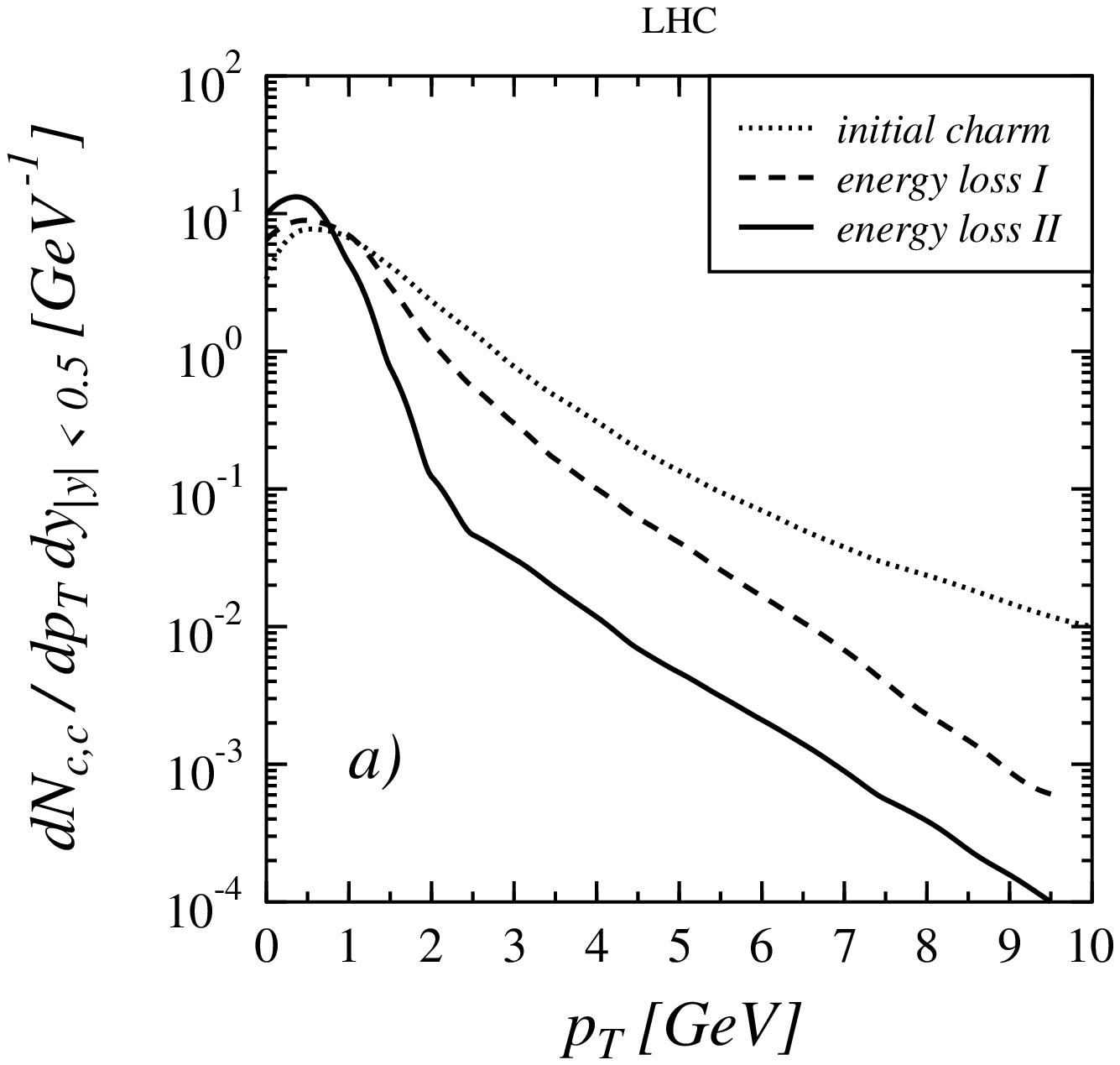}}
\centerline{\epsfxsize=.7 \hsize \epsffile{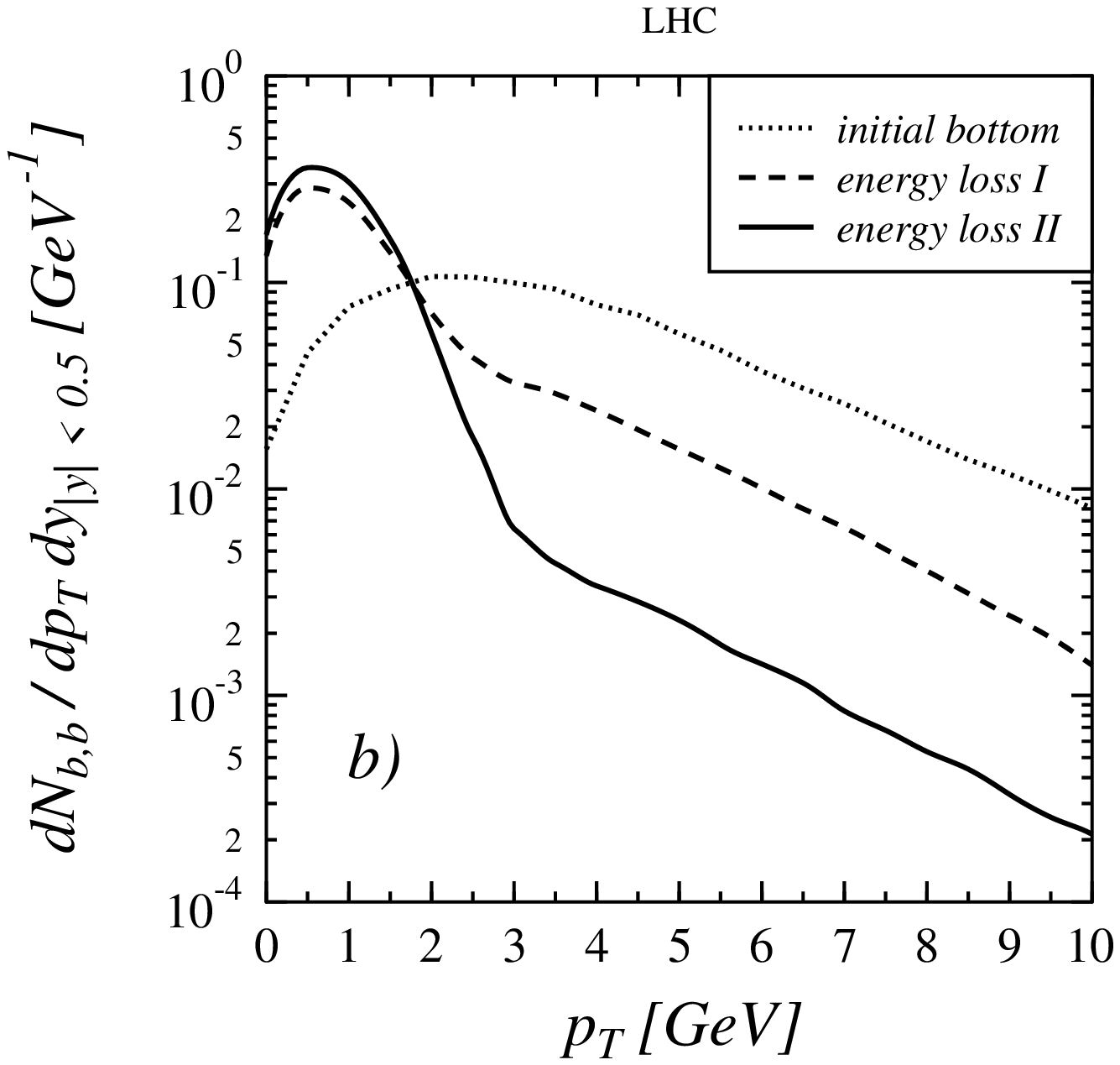}}
~\\[.1cm]
\caption{ {} }
\label{fig1}
\end{figure}

\newpage

\begin{figure}[h]
\centering
~\\[.1cm]
\centerline{\epsfxsize=.7 \hsize \epsffile{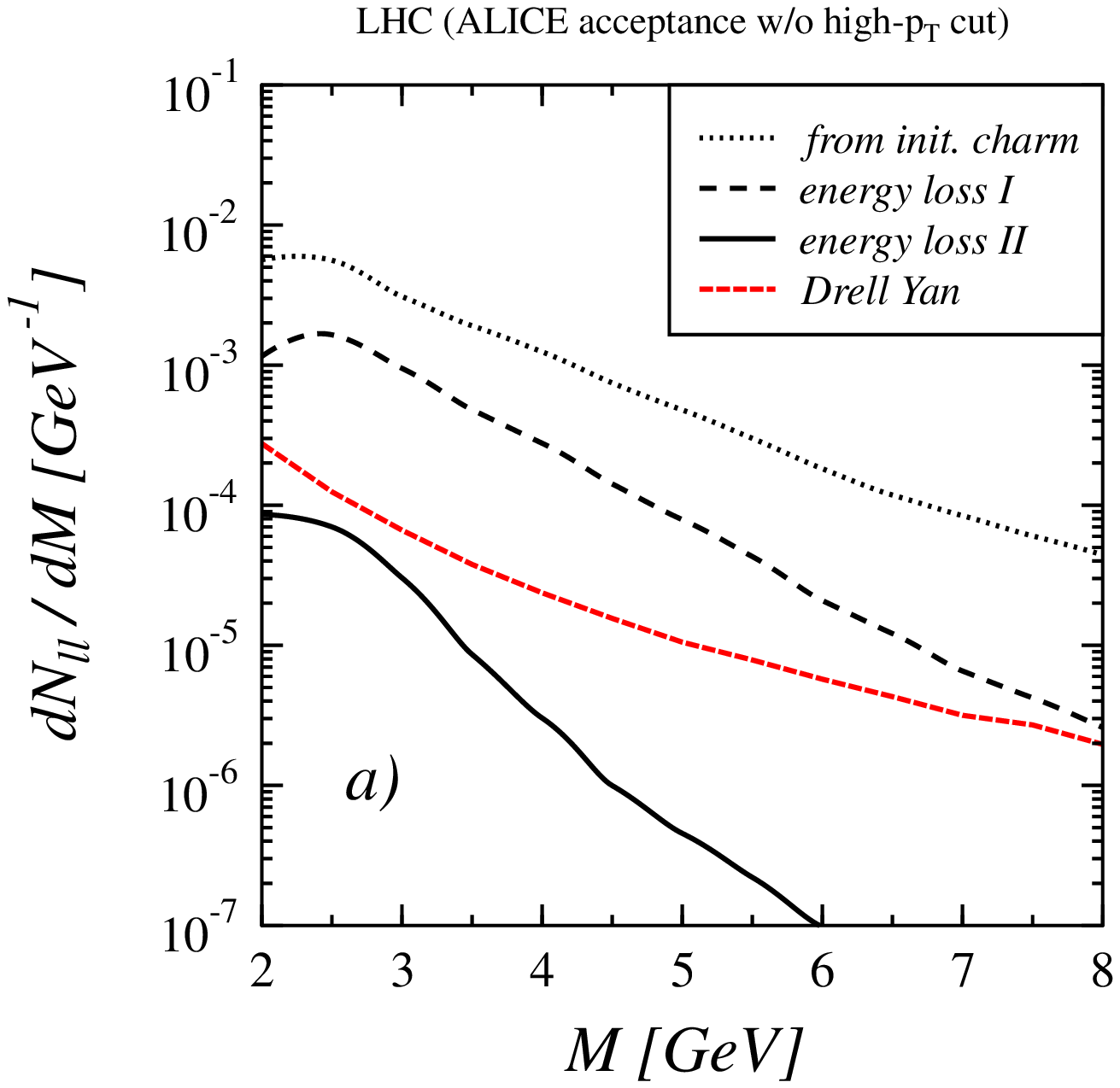}}
\centerline{\epsfxsize=.7 \hsize \epsffile{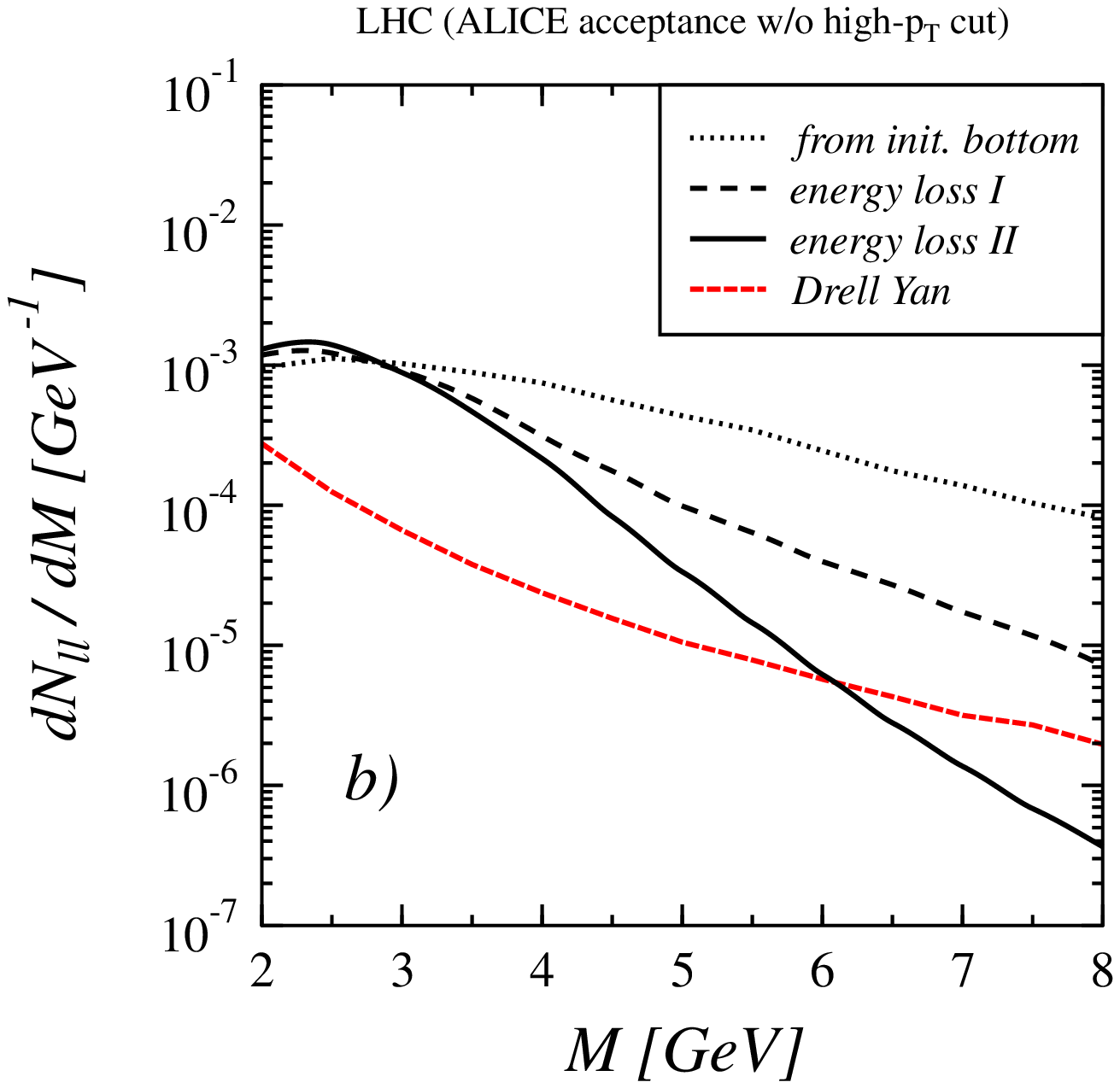}}
~\\[.1cm]
\caption{ {} }
\label{fig2}
\end{figure}

\begin{figure}[h]
\centering
~\\[.1cm]
\centerline{\epsfxsize=.8 \hsize \epsffile{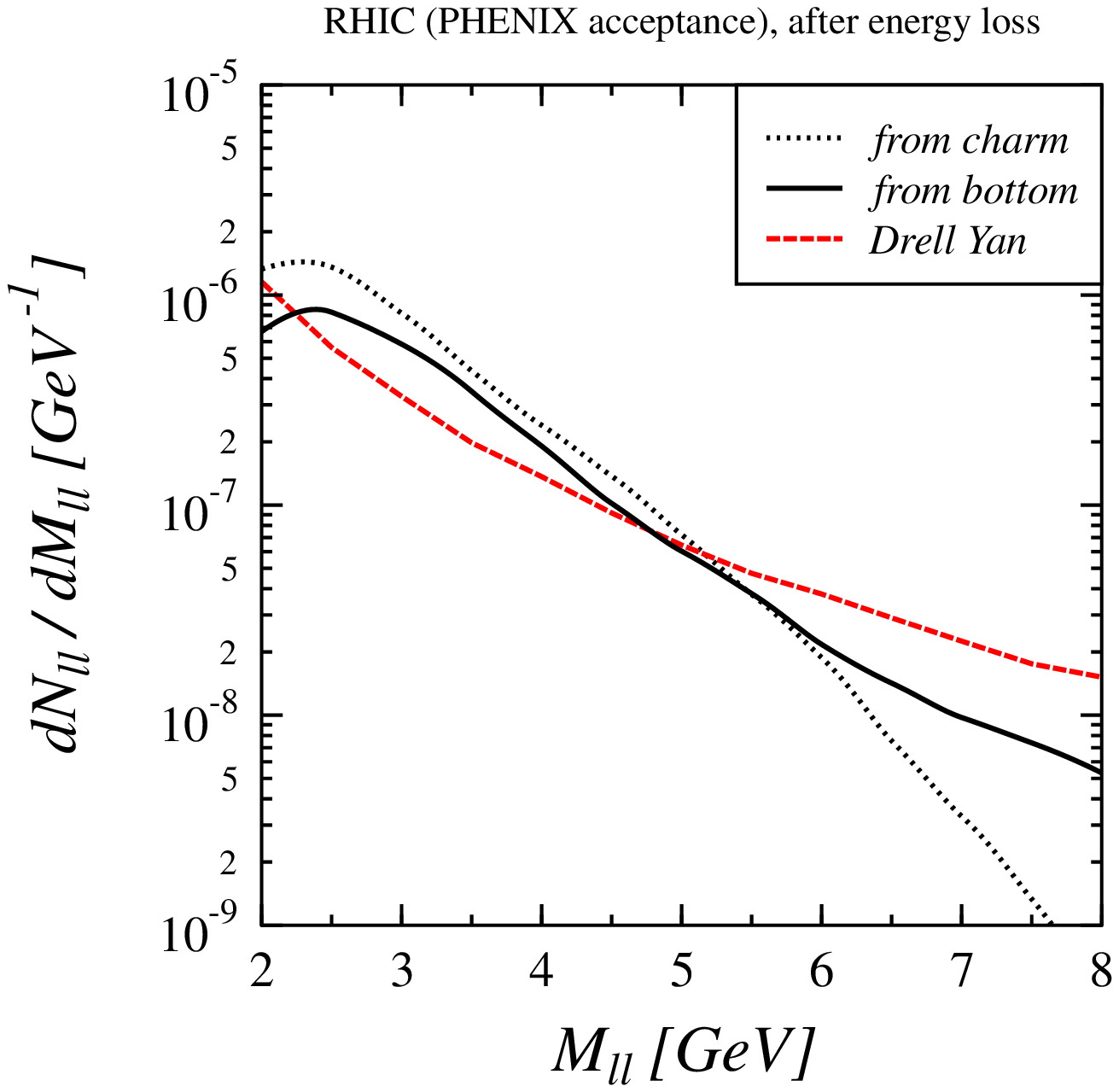}}
~\\[.1cm]
\caption{ {} }
\label{fig3}
\end{figure}

\end{document}